\documentstyle[a4,11pt,epsf,twoside,graphicx]{article}
\def\TeV{\ifmmode {\,\mathrm{ Te\kern -0.1em V}}\else
                   \textrm{Te\kern -0.1em V}\fi}%
\def\GeV{\ifmmode {\,\mathrm{ Ge\kern -0.1em V}}\else
                   \textrm{Ge\kern -0.1em V}\fi}%
\def\MeV{\ifmmode {\,\mathrm{ Me\kern -0.1em V}}\else
                   \textrm{Me\kern -0.1em V}\fi}%
\def\keV{\ifmmode {\,\mathrm{ ke\kern -0.1em V}}\else
                   \textrm{ke\kern -0.1em V}\fi}%
\def\eV{\ifmmode  {\,\mathrm{ e\kern -0.1em V}}\else
                   \textrm{e\kern -0.1em V}\fi}%

\newcommand{\fbi}{\,\rm{fb}^{-1}}

\newcommand{\WW}    {\mathrm{W}^+\mathrm{W}^-}
\newcommand{\ee}    {\mathrm{e}^+\mathrm{e}^-}

\begin{document}
\begin{titlepage}
\vspace*{-3cm}
\begin{flushright}
23rd October 2002
\end{flushright}
\vspace{4cm}
\begin{center}
{\huge \bf \boldmath Measurement of Trilinear Gauge Couplings
  at a $\gamma \gamma$ and $e \gamma$ Collider}

{\Large I. Bo\v zovi\'c-Jelisav\v ci\'c, K. M\"onig, J. \v Sekari\'c}

DESY-Zeuthen
\end{center}

\begin{abstract}
The processes $\gamma \gamma \rightarrow \WW$ and $e \gamma \rightarrow \nu W$
are sensitive to triple gauge boson interactions. 
Both reactions have been simulated
for hadronically decaying W-bosons and the sensitivity to anomalous couplings
has been estimated.
\end{abstract}
\vfill
Talk presented at LCWS2002, August 2002, 
   Jeju Island , Korea 
\end{titlepage}

\title{MEASUREMENT OF TRILINEAR GAUGE COUPLINGS
  AT A $\gamma \gamma$ AND $e \gamma$ COLLIDER
} 
\author{I. Bo\v zovi\'c-Jelisav\v ci\'c, K. M\"onig, J. \v Sekari\'c
\\ 
\\ 
        {\it DESY-Zeuthen}
}
\date{}
\maketitle
\begin{abstract}
The processes $\gamma \gamma \rightarrow \WW$ and $e \gamma \rightarrow \nu W$
are sensitive to triple gauge boson interactions. 
Both reactions have been simulated
for hadronically decaying W-bosons and the sensitivity to anomalous couplings
has been estimated.
\end{abstract}

\section{Introduction}
If no light Higgs boson exists weak interactions become strong at high energies
and for example WW scattering finally violates unitarity at 
$\sqrt{s} = 1.2 \TeV$.
First signs of this effect should be seen in a modification of the triple
gauge-boson couplings already at lower energies. 
If, on the contrary, the model remains perturbative up to very high energies
triple gauge couplings are modified by loop corrections making them
sensitive to new physics at high energy in the same way as the weak
mixing angle $\sin^2 \theta$, measured at LEP and SLD,
is sensitive to the Higgs boson mass.

\begin{figure}[htb]
\begin{center}
\includegraphics[width=0.3\linewidth]{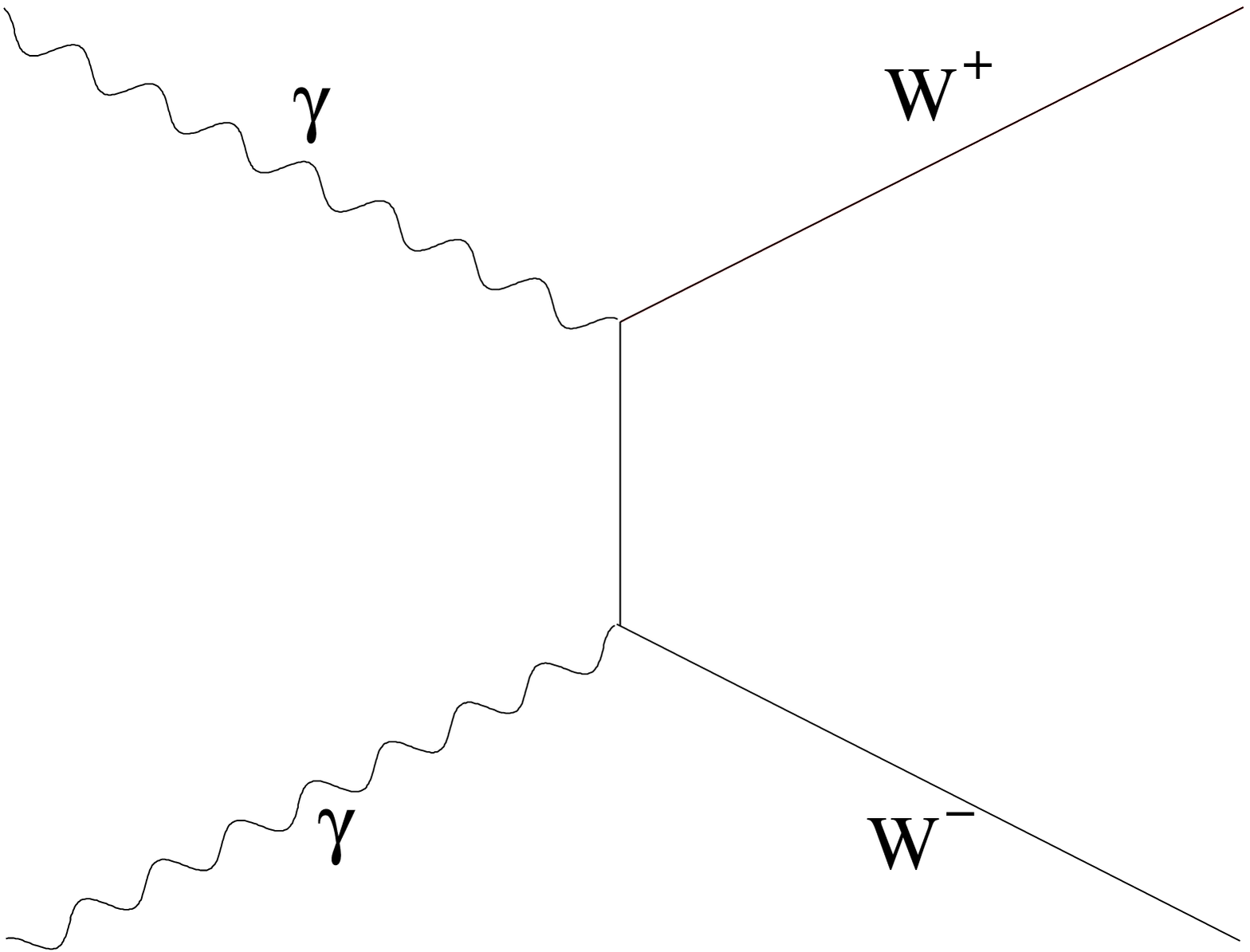}
\hspace{1cm}
\includegraphics[width=0.3\linewidth]{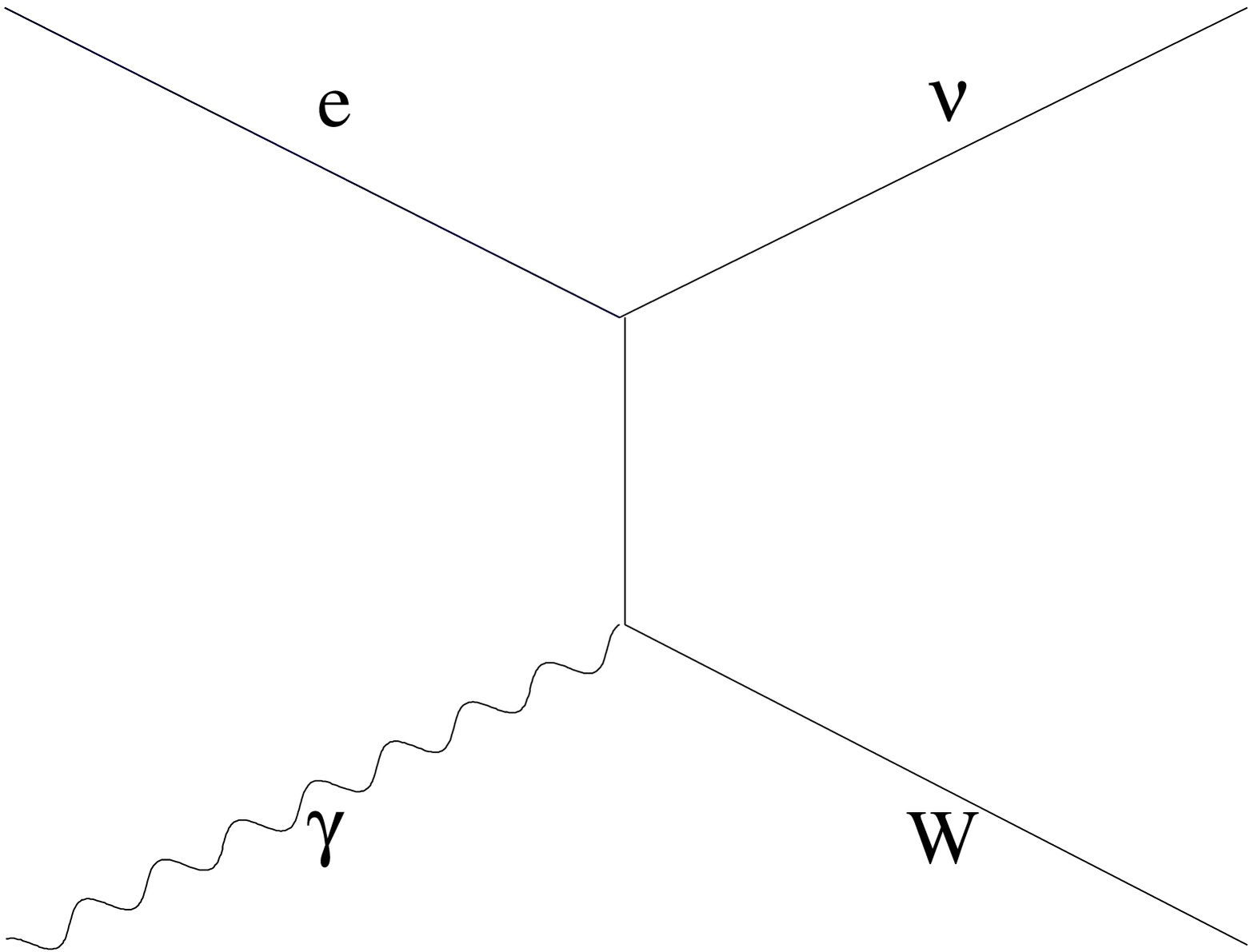}
\end{center}
\caption{Dominating Feynman diagrams for $\gamma \gamma \rightarrow \WW$
  and $e \gamma \rightarrow W \nu$
}
\label{fig:feyn}
\end{figure}

The cross sections 
 $\gamma \gamma \rightarrow \WW$ and $e \gamma \rightarrow \nu W$ are
about an order of magnitude larger than 
$ \ee \rightarrow \WW$ and the dominating Feynman diagrams,
shown in figure \ref{fig:feyn}, contain the
triple gauge coupling. Also in $\ee$ the WW$\gamma$ and WWZ couplings are
mixed. They can, however, be separated, if beam polarization is available.

A $\gamma \gamma$ collider can be realized if a linear collider is run in the
$e^- e^-$ mode and a high power laser beam is collided with the electron beam
a few mm in front of the interaction point. By Compton scattering photons of
up to 80\% of the beam energy can be obtained and the luminosity in the
high energy part of the spectrum can be as high as a third of the $\ee$
luminosity.
For $e \gamma$ one can either switch off one laser or one can use the
parasitic $e \gamma$ luminosity in the $\gamma \gamma$ mode using unconverted 
electrons. In the latter case it is, however, not possible to distinguish from
which of the two directions the electron was coming and one has additional
background from $\gamma \gamma$ collisions.

In this note the processes 
$\gamma \gamma \rightarrow \WW$ and $e \gamma \rightarrow \nu W$ are analyzed
requiring hadronic decays of the Ws. This ensures that the events can be fully
reconstructed. In the $\gamma \gamma$ fully hadronic mode in general the
$W^+$ and the $W^-$ cannot be distinguished. This is, however, no problem
since the cross section has to be symmetric in the production angle
$\cos \theta$.
Also for hadronically decaying Ws it is not possible in general to
distinguish the quark from the antiquark. This leaves a twofold
ambiguity in the decay angles. One can still separate longitudinally from
transversely polarized Ws, but the two transverse states cannot be separated.

$\gamma \gamma$ and $e \gamma$ colliders allow a high degree of beam 
polarization. For $\gamma \gamma$ one has two distinct polarization states
$J_z=0$ and $J_z=2$. For $e \gamma$ the electron has to be left-handed in
order to couple to W-bosons. The cross section then depends on the 
helicity of the photon, $J_\gamma$.

\begin{figure}[htb]
\begin{center}
\includegraphics[width=0.4\linewidth]{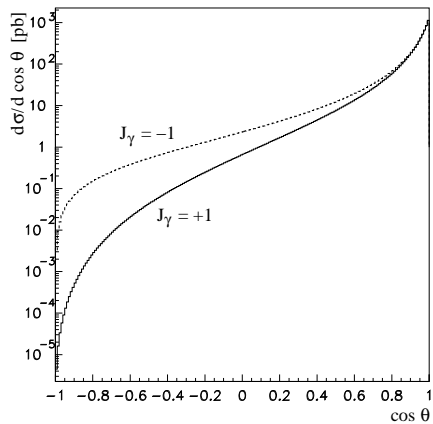}
\end{center}
\caption{Differential cross section 
$d \sigma(e \gamma \rightarrow \nu W) / d \cos \theta$ for the two $\gamma$
helicities.}
\label{fig:difsigg}
\end{figure}

\begin{figure}[htb]
\begin{center}
\includegraphics[width=0.4\linewidth]{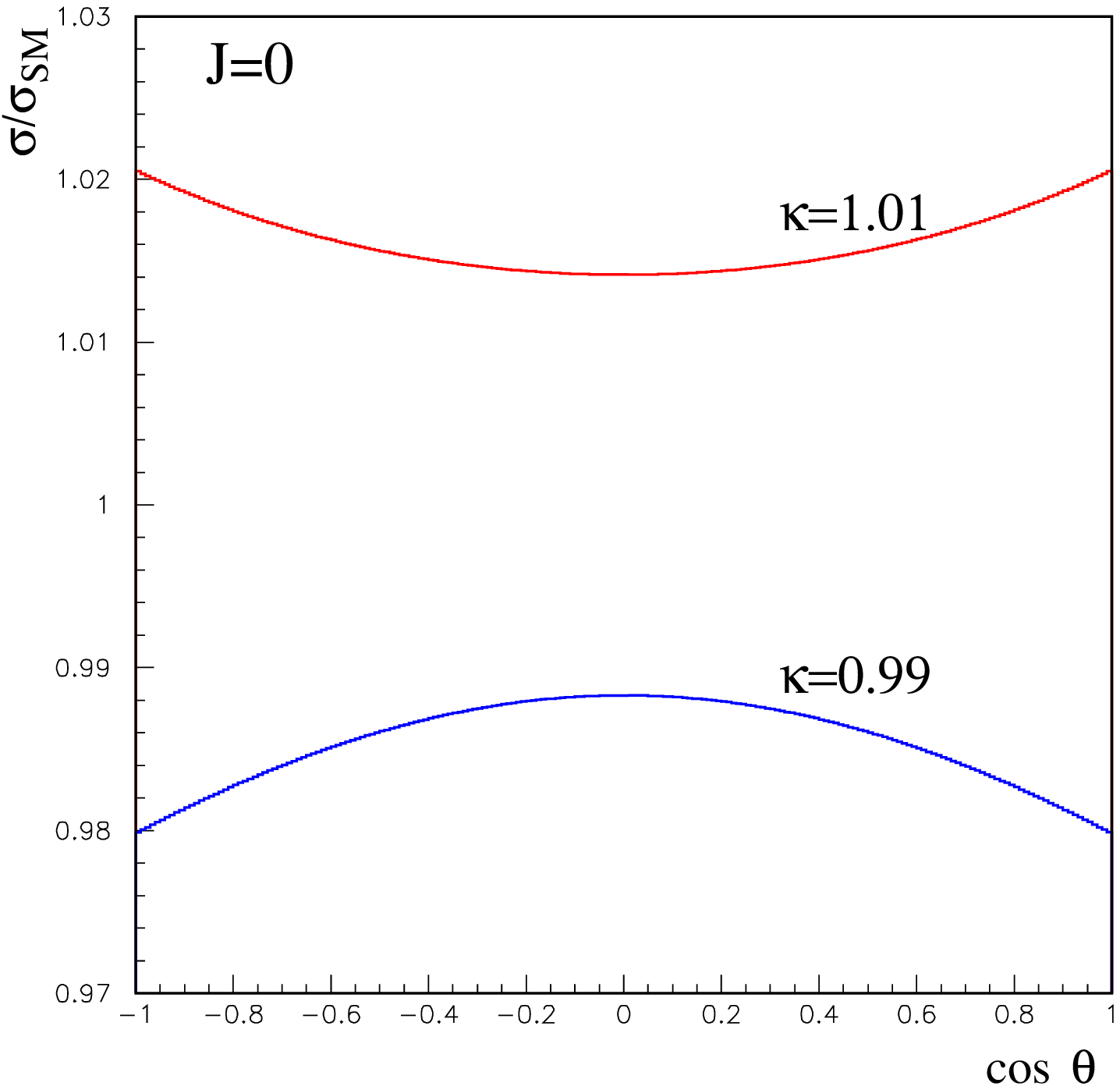}
\includegraphics[width=0.4\linewidth]{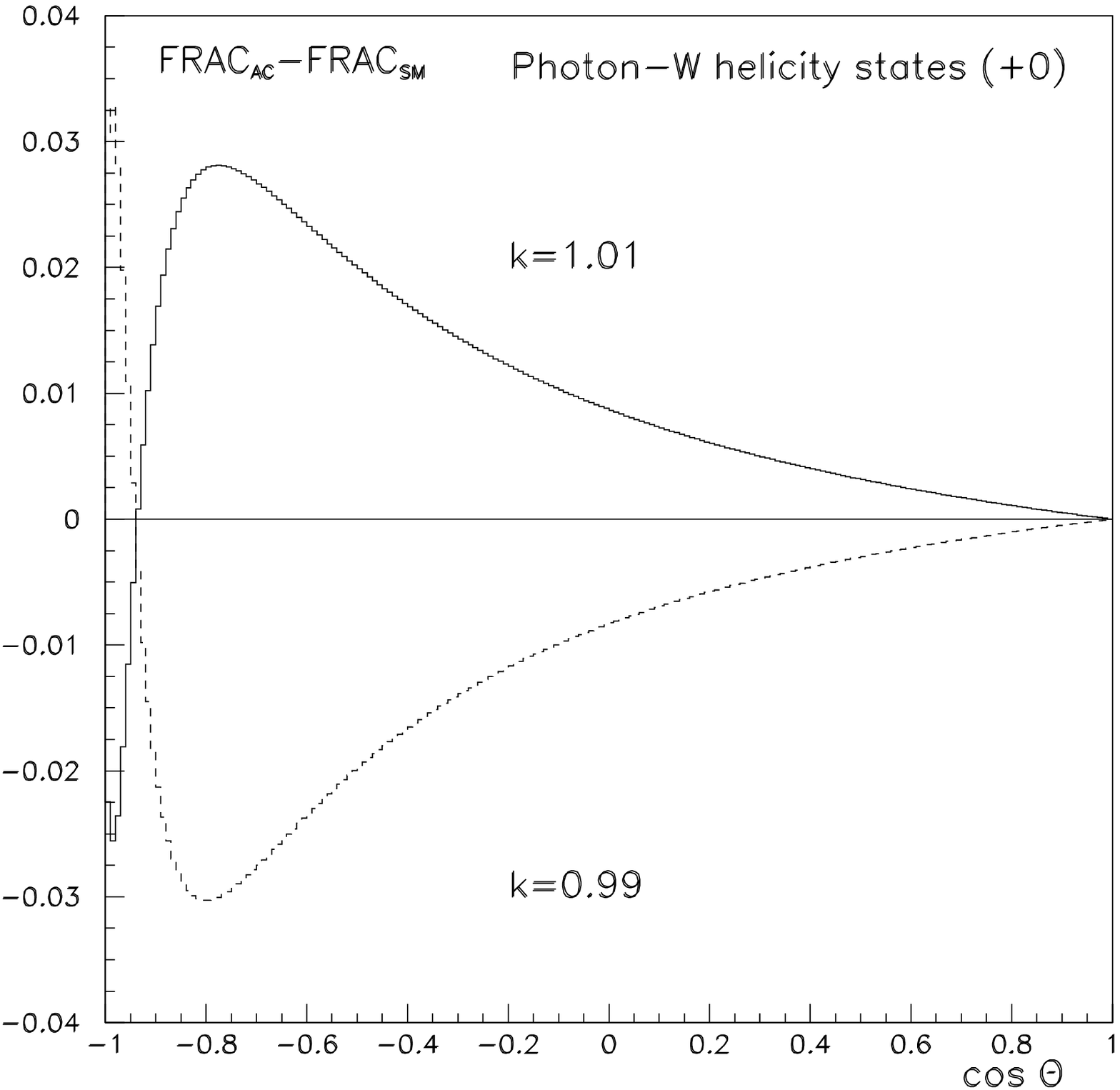}
\end{center}
\caption{a) Variation of the differential cross section 
$d \sigma(e \gamma \rightarrow \nu W) / d \cos \theta$ with the
anomalous coupling $\kappa_\gamma$. b) variation of the fraction of
longitudinal Ws in $e \gamma \rightarrow W \nu$ as a function of
the W production angle}
\label{fig:dsigcoup}
\end{figure}

Both processes proceed via t-channel exchange of a W and are strongly
forward peaked, as can be seen from figure \ref{fig:difsigg} for the
$e \gamma$ case. The anomalous couplings, which are parameterized with the
usual parameters $\kappa_\gamma,\, \lambda_\gamma$ \cite{ref:couppar}
manifest themselves with modifications of the total and differential cross 
section and of the polarization structure as seen in figure \ref{fig:dsigcoup}.

\section{Event Selection}
\begin{figure}[htb]
\begin{center}
\includegraphics[width=0.8\linewidth]{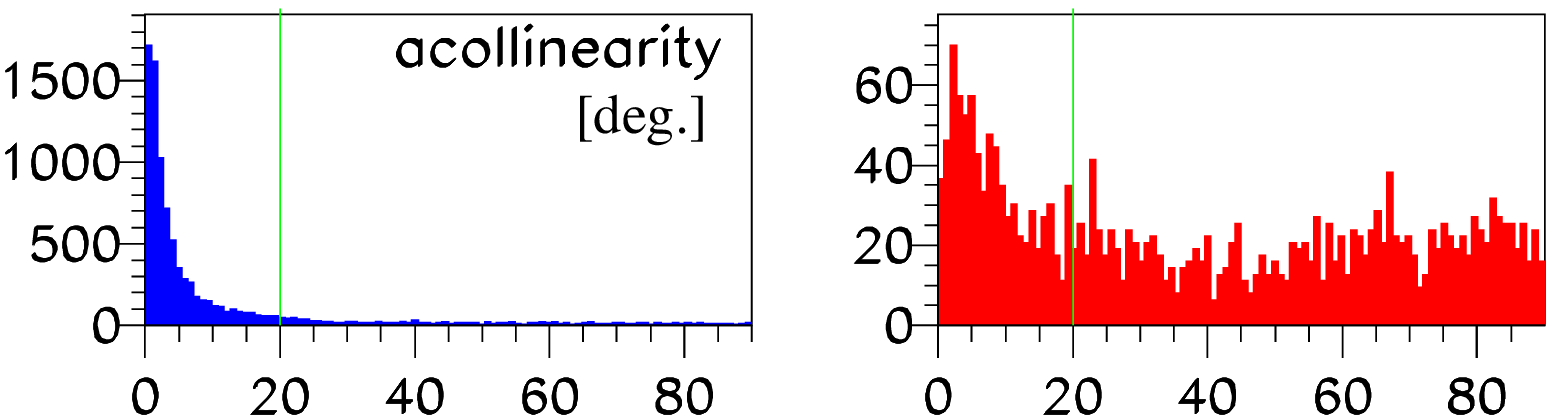}
\includegraphics[width=0.83\linewidth]{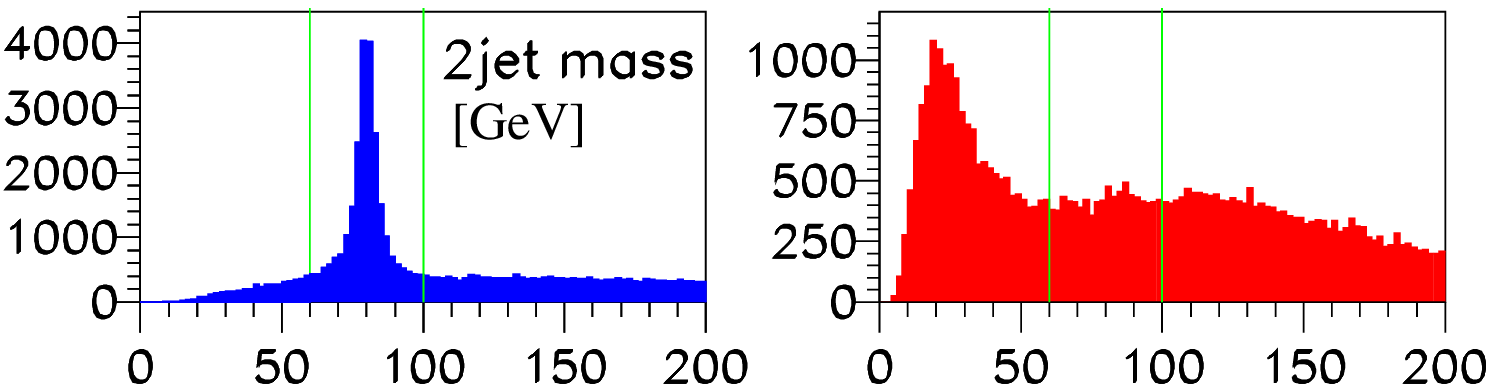}
\end{center}
\caption{Acolinearity and 2-jet mass for signal (left) and
background (right) in $\gamma \gamma \rightarrow \WW$}
\label{fig:accbg}
\end{figure}

The processes $\gamma \gamma \rightarrow \WW$ and $e \gamma \rightarrow \nu W$ 
and the relevant background processes have been simulated with
PYTHIA \cite{pythia} using the correct $\gamma$ energy spectra. 
The detector has been simulated with SIMDET \cite{simdet}, assuming in 
addition that no particles below $7^\circ$ are accepted.
For  $\gamma \gamma$ the only background is 
$\gamma \gamma \rightarrow q \bar{q}$.
For $e \gamma$ the relevant backgrounds are 
$e \gamma \rightarrow eZ$ and $e \gamma \rightarrow e q \bar{q}$. The
additional
backgrounds to $e \gamma \rightarrow \nu W$ in the $\gamma \gamma$ mode have 
not yet been studied. As shown for two examples in figure \ref{fig:accbg}
the backgrounds can be rejected efficiently with kinematic cuts.
In $\gamma \gamma$ a signal/background ratio of 15 for $J_z=2$ can be reached.
For $J_z=0$ it is much higher. In $e \gamma$ only some background is left at
extreme polar angles that can easily be cut without a significant loss in
the efficiency.
The efficiency for both processes is around 90\% over most of the solid angle
with a drop to about 80\% in the very forward region. In the forward region
the Pythia efficiencies might be somewhat optimistic since the polarization
structure of the Ws is not included. However in the final analysis a 
multi-dimensional acceptance function is used binned in the production angle
and the polar decay angles of the Ws. In this way only some small effects
stemming from the azimuthal decay angles are lost. The acceptance functions
have been cross checked at a fixed beam energy with WHIZARD \cite{whizard}
that includes the full polarization structure.

\section{Sensitivity Estimate}

To estimate the sensitivity of the two processes to anomalous couplings
helicity amplitudes for stable Ws have been used \cite{helamp}. 
The center of mass energy has been taken to be $0.8 \sqrt{s}$ for
$\gamma \gamma$ and $0.9 \sqrt{s}$ for $e \gamma$ which corresponds to the
maximum energy for these modes. The beam polarization has been assumed to
be 100\%. The events have been binned in the production angle and the
polar decay angle of the one or two Ws. The azimuthal decay angles, which
are sensitive to the interference of the different helicity amplitudes
are neglected for the moment.
Using these amplitudes event rates are predicted without anomalous couplings
in the different bins
in the production and decay angles assuming a given luminosity. These
event rates are then multiplied by the acceptance function to obtain
rates of measured events which are taken as ``data''. These ``data'' are
fitted with a $\chi^2$ fit using the same helicity amplitudes multiplied
by the acceptance function allowing for anomalous couplings. To allow
for a possible normalization uncertainty from the luminosity and
efficiency calculation, a free normalization factor has been included
in the fit which has been constrained with the assumed accuracy on the
normalization. The fits have been repeated with the normalization
fixed to unity.

Table \ref{tab:res} shows the results for $\sqrt{s}_{ee} = 500 \GeV,\,
{\cal L} = 110 \fbi$ and $J_z=2$ for $\gamma \gamma$ and $J_\gamma = 1$ for
$e \gamma$. For the results shown, only one anomalous coupling has been left
free in the fit while the other has been fixed to its Standard Model value.
In a fit with both couplings left free the correlations for $e \gamma$ are
small so that the errors practically don't change. For $\gamma \gamma$ the 
correlations are large and the errors increase by a factor four.
For the not shown polarization states the errors are up to a factor four 
larger than for the shown ones, where the difference increase with larger
normalization uncertainty.
Between $\sqrt{s}_{ee} = 500 \GeV$ and $\sqrt{s}_{ee} = 800 \GeV$ the 
sensitivity dependence on the center of mass energy is very small.

The $e \gamma$ results are given for the dedicated running mode. In the
parasitic mode during $\gamma\gamma$ running, the same sensitivity can
be reached for $\kappa_\gamma$ while the expected uncertainty for
$\lambda_\gamma$ is a factor two larger, if the same efficiency and
purity can be reached.

\begin{table}[htb]
\begin{center}
\begin{tabular}[c]{|c|c|c|c|c|c|c|}
\hline
 & \multicolumn{3}{c|}{$\gamma \gamma, \, J_z = 2$}
 & \multicolumn{3}{c|}{$e \gamma, \, J_\gamma = 1$}\\
\hline
  $\Delta\mathcal{L}$&  $1\%$ & $0.1\%$ & accurate 
                     &  $1\%$ & $0.1\%$ & accurate\\
  \hline
  $\Delta\kappa_{\gamma}\cdot 10^{-4}$ & 8.5 & 6.7 & 4.2 
                                       & 34 & 10 & \phantom{1}5\\
  \hline
  $\Delta\lambda_{\gamma}\cdot 10^{-4}$ & 6.3 & 6.0 & 5.0 
                                        & 16 & 15 & 15\\
  \hline  
\end{tabular}
\end{center}
\caption{Estimated sensitivity to $\kappa_\gamma$ and $\lambda_\gamma$
in $\gamma \gamma$ and $e \gamma$ with $\sqrt{s}_{ee}=500\GeV$ and
${\cal L} = 110 \fbi$.
}
\label{tab:res}
\end{table}

The sensitivity estimates don't include yet the azimuthal
decay angles which are sensitive to the interference of the helicity
amplitudes. Some improvement can be expected from the inclusion of these
angles. On the contrary, apart from the normalization no systematic 
uncertainties have been included up to now. Both studies are planned in the 
near future.

\section{Conclusion}

According to preliminary estimates the WW$\gamma$ triple gauge couplings
can be measured in $\gamma \gamma$ end $e \gamma$ collisions with
an accuracy about a factor two worse than in $\ee$ \cite{wolfgang}.
However the measurements in $\gamma \gamma$ end $e \gamma$ are in the
spacelike region at lower scales, so that they are complementary to the 
$\ee$ measurements, in case deviations from the Standard Model
predictions are found.

\end{document}